\documentclass[aps,prl,reprint,superscriptaddress,amsmath,floatfix,amssymb]{revtex4-2}

\usepackage{graphicx}
\usepackage{bm}
\usepackage{epstopdf}

\begin{document}

\title{``Ultra-weak'' First-Order Phase Transition in Biaxial Liquid Crystals}

\author{Soumyajit Pramanick}
\affiliation{Department of Physics, Lady Brabourne College, Kolkata 700017, India}
\author{Mrinal Kanti Debnath}
\affiliation{Department of Physics, Ramsaday College, Amta, Howrah, West Bengal, India}
\author{Nababrata Ghoshal}
\affiliation{Department of Physics, Ramsaday College, Amta, Howrah, West Bengal, India}
\author{Soumen Kumar Roy}
\affiliation{Department of Physics, Jadavpur University (Rtd), Calcutta 700032, India}
\author{Sudeshna Dasgupta}
\affiliation{Department of Physics, Lady Brabourne College, Kolkata 700017, India}

\date{\today}

\begin{abstract}
We report high-resolution Monte Carlo evidence characterizing the hierarchy of transition strengths in biaxial nematogens. By employing multiple histogram reweighting on large lattices, we demonstrate that while the isotropic-to-uniaxial ($I \to N_U$) transition is symmetry-enforced first-order, the uniaxial-to-biaxial ($N_U \to N_B$) transition in the region between the tricritical point and the triple point ($\lambda = 0.26$) is proximity-driven and more than an order of magnitude smaller for large system sizes. Such a class of ultra-weak first-order transitions could be a new pathway for future investigations.
\end{abstract}

\maketitle

The orientational transition in the three-dimensional Lebwohl--Lasher (LL) ~\cite{Lebwohl} model is historically celebrated as the archetype of a weak first-order transition ~\cite{Zhang1992}. In this Letter, we demonstrate that the uniaxial-to-biaxial ($N_U \to N_B$) transition in biaxial nematogenic systems described by Straley's quadrupolar potential reveals a remarkably more delicate, ``ultra-weak'' first-order regime. At a molecular biaxiality of $\lambda = 0.26$, the proximity to the tricritical point ($\lambda_t \approx 0.203$) suppresses the bulk free-energy barrier of this secondary transition to a level orders of magnitude smaller for large systems than that of the already weak isotropic-to-uniaxial benchmark, presenting an extraordinary challenge for conventional numerical and experimental resolution. Fundamental to our understanding of such systems is the nature of the phase transitions themselves, particularly the difference between symmetry-enforced first-order transitions and proximity-enforced ones. Halperin, Lubensky, and Ma predicted that transitions traditionally viewed as second-order, such as the superconducting and smectic-A transitions, are driven ``weakly'' first-order by the coupling of the complex order parameter to gauge-field fluctuations ~\cite{Halperin1974}. In these systems, taking the trace over the configurations of the vector potential generates an effective free energy containing a negative cubic-like term proportional to $-|\psi|^3$, which inevitably forces a discontinuous jump ~\cite{Halperin1974}. 

The biaxial nematogenic systems described by Straley's quadrupolar potential ~\cite{Straley1974} exhibit a similar but more nuanced hierarchy of transition strengths. We focus on the regime where molecular biaxiality $\lambda$ is tuned between the tricritical point ($\lambda_t \approx 0.203$) and the triple point ($\lambda_c \approx 0.27$) ~\cite{Debnath2025}. To quantify the ``strength'' of these transitions, we examine the Landau free-energy density expansions permitted by the symmetry reduction of each phase. 

The transition from the isotropic melt to the uniaxial nematic phase ($I \to N_U$) involves an $O(3) \to D_{\infty h}$ symmetry reduction. The expansion of the free energy $F_U$ in terms of the uniaxial order parameter $S$ is:
\begin{equation} 
F_U = F_I + \frac{1}{2} a_u S^2 - \frac{1}{3} b_u S^3 + \frac{1}{4} c_u S^4
\end{equation}
where the cubic term ($S^3$) is a fundamental requirement of the orientational geometry ~\cite{Chaikin1995}. This term enforces a relatively robust first-order jump, creating a well-defined free-energy barrier $\Delta A$ that we resolve clearly in our simulations.

Conversely, for the transition from the uniaxial to the biaxial nematic phase ($N_U \to N_B$), the $D_{\infty h}$ symmetry of the reference uniaxial phase forbids a cubic (or any odd order terms in general) term in the biaxial order parameter $B$. The relevant expansion is:
\begin{equation}
F_B = F_{U} + \frac{1}{2} a_b B^2 + \frac{1}{4} \tilde{c}_b B^4 + \frac{1}{6} e_b B^6
\end{equation}
When $\tilde{c}_b$ is positive, the sixth-order term can be neglected and the transition is of second order.  At $\lambda = 0.26$, the system is in a first-order regime because the effective quartic coefficient $\tilde{c}_b$ has become negative close to the tricritical point, and the sixth-order term is required for the stability of the system ~\cite{Chaikin1995}. However, because this discontinuity is proximity-driven—with $\tilde{c}_b$ only slightly below zero as it approaches the tricritical point—the resulting jump is ``ultra-weak.''

We model the system using a simple-cubic lattice of $N = L^3$ sites, where each site hosts a biaxial molecule defined by three orthogonal unit vectors $\{\mathbf{e}, \mathbf{e}_{\perp}, \mathbf{m}\}$. The molecules interact with their nearest neighbors via Straley's quadrupolar pair potential ~\cite{Straley1974}. In the Sonnet--Virga--Durand (SVD) parameterization ~\cite{Sonnet2003}, the uniaxial-biaxial coupling parameter $\gamma$ is set to zero, yielding the potential:
\begin{equation}
\begin{split}
V = -\epsilon \{P_2(\mathbf{m}_i \cdot \mathbf{m}_j) +  \lambda [2(P_2(\mathbf{e}_i \cdot \mathbf{e}_j) \\ + P_2(\mathbf{e}_{\perp i} \cdot \mathbf{e}_{\perp j})) - P_2(\mathbf{m}_i \cdot \mathbf{m}_j)]\}
\end{split}
\end{equation}
where $\epsilon$ defines the energy scale and $\lambda$ is the molecular biaxiality parameter ~\cite{Chiccoli1999}. This potential is equivalent to a symmetry-adapted expansion in $D_{2h}$ Wigner functions:
\begin{equation}
V = -\epsilon \{ R^2_{00}(\Omega_{ij}) + 6\lambda R^2_{22}(\Omega_{ij}) \}
\end{equation}
where $\langle R^2_{00} \rangle$ and $\langle R^2_{22} \rangle$ serve as the primary uniaxial ($S$) and biaxial ($C$) order parameters, respectively ~\cite{Biscarini1995}.

To resolve the thermodynamic order and transition strengths, we utilize the multiple histogram reweighting (MHR) technique to construct a high-precision free-energy-like function $A(E) = -\ln P(E)$ ~\cite{Challa1986,Swendsen1987}. For first-order transitions, $A(E)$ exhibits a double-well structure. We quantify the transition strength by the barrier height $\Delta A = A(E_m) - A(E_1)$, where $E_m$ is the local maximum and $E_1$ represents either minimum of equal depth. Our simulations on lattices up to $L^3 = 60^3$ allow for the definitive characterization of the order-of-magnitude difference in $\Delta A$ between the orientational condensation of the long axes and the subsequent ordering of the transverse axes.

We have performed extensive Monte Carlo simulations at a molecular biaxiality of $\lambda = 0.26$. This specific parameter value is situated in the narrow thermodynamic window between the tricritical point ($\lambda_t \approx 0.203$) and the triple point ($\lambda_c \approx 0.27$) identified in our previous numerical work ~\cite{Debnath2025}. In this regime, with lowering temperature, the system undergoes a sequential symmetry reduction: first from the isotropic ($I$) phase to the uniaxial nematic ($N_U$) phase, followed by a second transition to the biaxial nematic ($N_B$) phase.

The temperature evolution of the primary uniaxial ($\langle R^2_{00} \rangle$) and biaxial ($\langle R^2_{22} \rangle$) order parameters across four system sizes is presented in Fig.~1. At $\lambda = 0.26$, the two transitions are nearly degenerate in temperature. For the smaller lattices ($L=20, 30$), finite-size fluctuations dominate the system, causing the onset of order to appear relatively smooth and masking the discontinuous nature of the transitions. However, as the system size increases to $L=60$, the curves sharpen significantly, revealing a distinct, albeit narrow, plateau of uniaxial stability before the biaxial order parameter begins its rapid ascent.

\begin{figure}[htbp]
\centering
\includegraphics[angle=-90, width=\columnwidth]{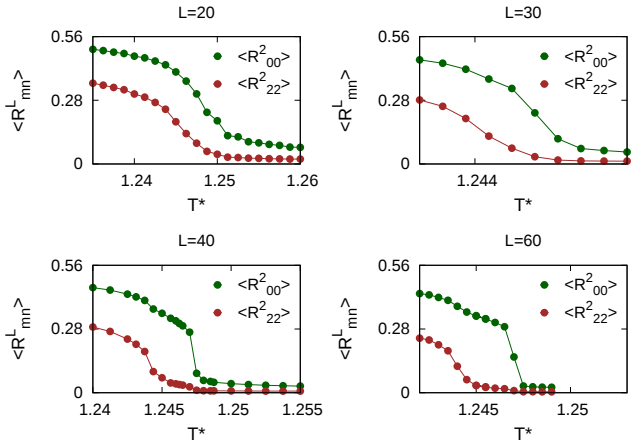}
\caption{Uniaxial and Biaxial order parameters $R^2_{00}$ and $R^2_{22}$ as a function of temperature for the isotropic-to-uniaxial ($I \to N_U$) and uniaxial-to-biaxial ($N_U \to N_B$) transitions at $\lambda = 0.26$ for various system sizes.}
\label{fig:order_parameters}
\end{figure}

This sequence is further clarified by the configurational specific heat, $C_V$ (Fig.~2). For $L=20$, the thermal response is dominated by a single, broad peak and we barely resolve the two transitions. As we move to larger system sizes, our simulations successfully resolve dual peaks. The high-temperature peak corresponds to the $I \to N_U$ orientational condensation, while the lower-temperature peak signifies the $N_U \to N_B$ ordering of the transverse axes. The sharpening of these peaks with $L$ is a hallmark of first-order behavior in this region of the phase diagram ~\cite{Challa1986,Zhang1992}.

\begin{figure}[htbp]
\centering
\includegraphics[angle=-90, width=\columnwidth]{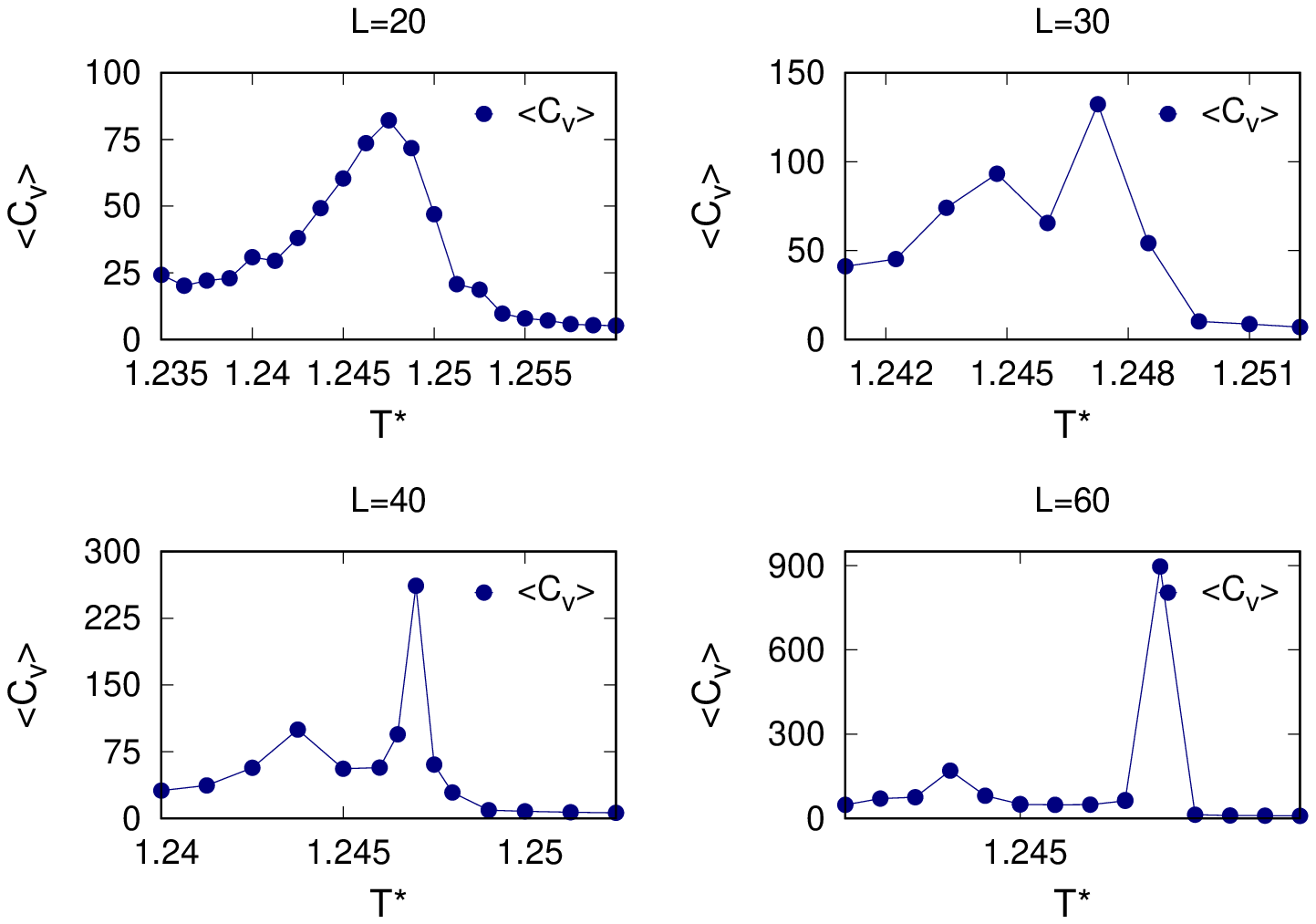}
\caption{$C_V$ as a function of temperature for the isotropic-to-uniaxial ($I \to N_U$) and uniaxial-to-biaxial ($N_U \to N_B$) transitions at $\lambda = 0.26$ for various system sizes.}
\label{fig:Cv}
\end{figure}

The central finding of this Letter lies in the quantitative comparison of the free-energy barriers. Using the multiple histogram reweighting (MHR) technique, we extracted the energy probability distributions $P(E)$ (Fig.~3) and the associated free-energy-like functions $A(E) = -\ln P(E)$ (Fig.~4). 

\begin{figure}[htbp]
\centering
\includegraphics[angle=-90, width=\columnwidth]{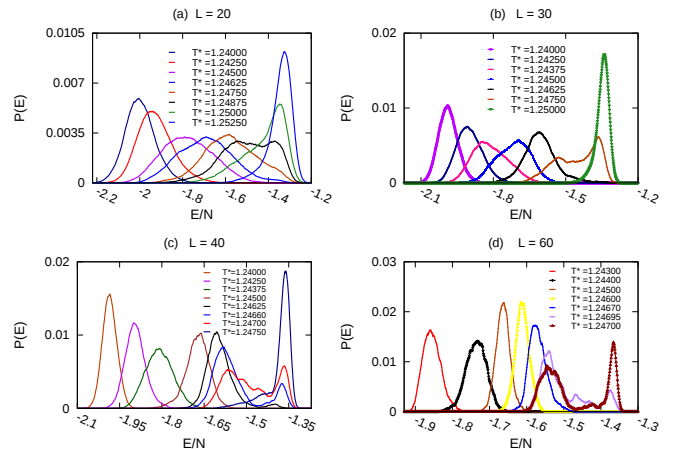}
\caption{Energy probability distributions $P(E)$ near the transition temperatures. The broadening and eventual peak-splitting at $L=60$ confirm the onset of phase coexistence.}
\label{fig:histograms}
\end{figure}

At the uniaxial-to-biaxial transition temperature $T^*_{UB}$, the system develops a double-well structure in $A(E)$, signaling phase coexistence. However, this barrier is remarkably fragile. As illustrated in Fig.~4, the barrier height $\Delta A$ for the $N_U \to N_B$ transition for large system sizes is more than an order of magnitude smaller than the corresponding barrier for the $I \to N_U$ transition at the same biaxiality (shown in the inset). Specifically, for our largest system ($L=60$), the $I \to N_U$ transition presents a robust barrier that is clearly symmetry-enforced by the cubic term in the Landau expansion ~\cite{deGennes1993}. In contrast, the $N_U \to N_B$ transition is proximity-driven; because $\lambda=0.26$ is close to the tricritical point where $\Delta A$ vanishes, the transition is ``ultra-weak''. We also note that finite-size effects at small $L$ can transiently invert the barrier ordering ($L=20$).

\begin{figure}[htbp]
\centering
\includegraphics[angle=-90, width=\columnwidth]{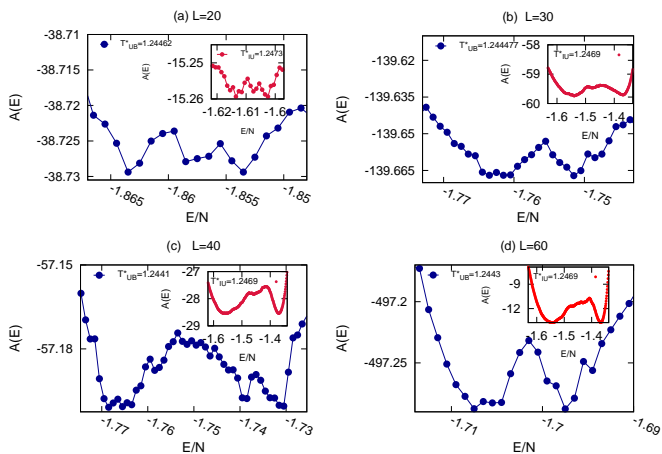}
\caption{Free-energy-like function $A(E)$ as a function of energy per particle for the uniaxial-to-biaxial ($N_U \to N_B$) transition at $\lambda = 0.26$. The inset shows the significantly larger barrier for the isotropic-to-uniaxial ($I \to N_U$) transition at the same biaxiality, highlighting the ``ultra-weak'' nature of the former.}
\label{fig:free_energy}
\end{figure}

\begin{table}[htbp]
\caption{Transition data for the isotropic-to-uniaxial ($I \to N_U$) and uniaxial-to-biaxial ($N_U \to N_B$) transitions at $\lambda = 0.26$.}
\begin{ruledtabular}
\begin{tabular}{cccccc}
System size & \multicolumn{2}{c}{Isotropic-Uniaxial} & \multicolumn{2}{c}{Uniaxial-Biaxial} \\
 & $T$ & $\Delta A$ & $T$ & $\Delta A$ \\
$20^3$ & 1.24729 & 0.005 & 1.24462 & 0.006 \\
$30^3$ & 1.24728 & 0.410 & 1.24448 & 0.015 \\
$40^3$ & 1.24692 & 0.999 & 1.24412 & 0.026 \\
$60^3$ & 1.24696 & 3.022 & 1.24430 & 0.056 \\
\end{tabular}
\end{ruledtabular}
\end{table}

Our large-scale Monte Carlo simulations establish the hierarchy of transition strengths in thermotropic biaxial nematogens under the SVD parameterization. By utilizing multiple histogram reweighting on lattices up to $L^3 = 60^3$, we have resolved the first-order nature of the $N_U \to N_B$ transition. We have demonstrated that, for large systems, the free-energy barrier for the uniaxial-to-biaxial transition at $\lambda = 0.26$ is more than an order of magnitude smaller, , than that of the already ``weak'' first-order $I \to N_U$ transition,approaching two orders of magnitude for $L=60$.

In conclusion, the mechanism identified by Halperin, Lubensky, and Ma ~\cite{Halperin1974}, whereby gauge-field fluctuations drive a second-order transition to weakly first-order by generating a negative cubic-like term, is analogous to the mechanism behind the relatively robust first-order nature of our $I \to N_U$ transition ~\cite{KamalaLatha2014} driven by a symmetry-enforced cubic term ($S^3$) in the Landau free-energy density expansion. However, the ``ultra-weak'' $N_U \to N_B$ transition we report here is different in origin. The $N_U \to N_B$ transition in the SVD model is proximity-driven. The $D_{\infty h}$ symmetry of the reference uniaxial phase forbids a cubic term for the biaxial order parameter, meaning that the transition is intrinsically continuous below the tricritical point $\lambda_t \approx 0.203$ ~\cite{DeMatteis2005}. The first-order character emerges only because the quartic coefficient $\tilde{c}_b$ is tuned to negative values by the molecular biaxiality parameter $\lambda$. Because our study focuses on the region just past this sign-flip, the resulting jump is significantly more delicate than the gauge-field-driven jumps described by HLM. We suggest that this class of ultra-weak first-order transitions could be universal in systems where cubic terms are forbidden by symmetry, forcing reliance on negative higher-order invariants, representing an avenue for future investigations.

\end{document}